\documentclass[manuscript,tighten]{aastex6}

\usepackage{soul}
\usepackage{datetime}

\usepackage{hyperref}
\hypersetup{colorlinks=true,urlcolor=blue}

\newcommand{\uhz}{\ensuremath{\mu{\rm Hz}}}

\newcommand{\ikt}{{\it Kepler}}
\newcommand{\ik}{{\it Kepler~}}

\shorttitle{K2 Neptune}
\shortauthors{Rowe et al.}
\slugcomment{Revision 4.4 --- \today --- \currenttime}

\begin{document}
\title{Time-Series Analysis of Broadband Photometry of Neptune from K2}

\author{Jason F. Rowe\altaffilmark{1}}
\affil{Institut de recherche sur les exoplan\`etes, iREx, D\'epartement de physique, Universit\'e de Montr\'eal, Montr\'eal, QC, H3C 3J7, Canada}
\altaffiltext{1}{jason@astro.umontreal.ca}

\author{Patrick Gaulme}
\affil{New Mexico State University, Department of Astronomy, P.O. Box 30001, Las Cruces, NM 88003-4500, USA}\
\affil{Apache Point Observatory, 2001 Apache Point Road, P.O. Box 59, Sunspot, NM 88349, USA}

\author{Jack J. Lissauer}
\affil{NASA Ames Research Center, Space Science \& Astrobiology Divison, MS 245-3, Moffett Field, CA 94035 USA}

\author{Mark S. Marley}
\affil{NASA Ames Research Center, Space Science \& Astrobiology Divison, MS 245-3, Moffett Field, CA 94035 USA}

\author{Amy A. Simon}
\affil{NASA Goddard Space Flight Center, Solar System Exploration Division (690.0), 8800 Greenbelt Road, Greenbelt, MD, 20771, USA}

\author{Heidi B. Hammel}
\affil{AURA, Inc., 1331 Pennsylvania Avenue NW, Suite 1475, Washington, DC 20004, USA}

\author{V\'ictor Silva Aguirre}
\affil{Stellar Astrophysics Centre, Department of Physics and Astronomy, Aarhus University, Ny Munkegade 120, DK-8000 Aarhus C, Denmark}

\author{Thomas Barclay}
\affil{Bay Area Environmental Research Institute, 625 2nd St. Ste 209 Petaluma, CA 94952, USA}

\author{Othman Benomar}
\affil{Center for Space Science, NYUAD Institute, New York University Abu Dhabi, PO Box 129188, Abu Dhabi, UAE}

\author{Patrick Boumier}
\affil{Institut d'Astrophysique Spatiale, CNRS and Universit\'e Paris Sud, (UMR 8617), F-91405 Orsay Cedex, France}

\author{Douglas A. Caldwell}
\affil{SETI Institute, Mountain View, CA 94043, USA}

\author{Sarah L. Casewell}
\affil{University of Leicester, Department of Physics and Astronomy, University Road, Leicester, LE1 7RH, UK}

\author{William J. Chaplin}
\affil{School of Physics \& Astronomy, University of Birmingham, Edgbaston, Birmingham, B15 2TT, UK}
\affil{Stellar Astrophysics Centre (SAC), Department of Physics and Astronomy, Aarhus University, Ny Munkegade 120, DK-8000 Aarhus C, Denmark}

\author{Knicole D. Col\'{o}n}
\affil{Bay Area Environmental Research Institute, 625 2nd St. Ste 209 Petaluma, CA 94952, USA}

\author{Enrico Corsaro}
\affil{Laboratoire AIM Paris-Saclay, CEA/DRF - CNRS - Universit\'e Paris Diderot, IRFU/SAp Centre de Saclay, F-91191 Gif-sur-Yvette Cedex, France}
\affil{Instituto de Astrof\'{i}sica de Canarias, E-38200 La Laguna, Tenerife, Spain}
\affil{Departamento de Astrof\'{i}sica, Universidad de La Laguna, E-38205 La Laguna, Tenerife, Spain}

\author{G.R.~Davies}
\affil{School of Physics and Astronomy, University of Birmingham, Birmingham, B15 2TT, United Kingdom}
\affil{Stellar Astrophysics Centre (SAC), Department of Physics and Astronomy,  Aarhus University, Ny Munkegade 120, DK-8000 Aarhus C, Denmark}

\author{Jonathan J. Fortney}
\affil{University of California, Santa Cruz, Department of Astronomy and Astrophysics, 1156 High Street, 275 Interdisciplinary Sciences Building, Santa Cruz, CA 95064, USA}

\author{Rafael A. Garcia}
\affil{Service d'Astrophysique, IRFU/DRF/CEA Saclay
L'Orme des Merisiers, bat. 709, 91191 Gif-sur-Yvette Cedex, France}

\author{John E. Gizis}
\affil{University of Delaware, Department of Physics and Astronomy, 104 The Green, Newark, DE 19716, USA}

\author{Michael R. Haas}
\affil{NASA Ames Research Center, Space Science \& Astrobiology Divison, MS 245-3, Moffett Field, CA 94035 USA}

\author{Beno\^{i}t Mosser}
\affil{LESIA, Observatoire de Paris, PSL Research University, CNRS, Universit\'e Pierre et Marie Curie, Universit\'e Paris Diderot, 92195 Meudon, France}

\author{Fran\c{c}ois-Xavier Schmider}
\affil{Laboratoire Lagrange, Observatoire de la C\^{o}te d'Azur, Universit\'e de Nice-Sophia-Antipolis, CNRS, Nice, France}

\begin{abstract}

\replaced{We present methods used to extract time-series photometry of Neptune from K2 observations and the subsequent correction of instrumental effects such as intrapixel variability and gain variations.  We present 49 days of nearly continuous broadband photometry of the planet Neptune.  The resultant time-series was analyzed to search for excess power that may be due to intrinsic global oscillations of the planet Neptune.  No evidence of global oscillations was found. We place an upper limit of $\sim$5 ppm at 1000 \uhz\ for the detection of a coherent signal.  With an observed cadence of 1-minute and point-to-point scatter less than 0.01\%, the photometric signal is dominated by reflected light from the Sun, which is in turn modulated by atmospheric variability of Neptune at the 2\% level.  A decrease in flux is also observed due to the increasing distance between Neptune and the K2 spacecraft, and solar variability with convection-driven solar p modes present.}  
{We report here on our search for excess power in photometry of Neptune collected by the K2 mission that may be due to intrinsic global oscillations of the planet Neptune. To conduct this search, we developed new methods to correct for instrumental effects such as intrapixel variability and gain variations. We then extracted and analyzed the time-series photometry of Neptune from 49 days of nearly continuous broadband photometry of the planet. We find no evidence of global oscillations and place an upper limit of $\sim$5 ppm at 1000 \uhz\ for the detection of a coherent signal.  With an observed cadence of 1-minute and point-to-point scatter less than 0.01\%, the photometric signal is dominated by reflected light from the Sun, which is in turn modulated by atmospheric variability of Neptune at the 2\% level.  A change in flux is also observed due to the increasing distance between Neptune and the K2 spacecraft, and solar variability with convection-driven solar p modes present.}

\end{abstract}


\section{Introduction}\label{intro}

The Solar System hosts two classes of giant planets: gas giants and ice giants. The ice giants are \replaced{smaller}{lower mass} (Uranus and Neptune masses are less than 18 Earth masses) with smaller hydrogen-helium envelopes, and exhibit strong enrichments in heavier elements.  The \ik Mission has aptly demonstrated that Neptune- and Uranus-sized planets are common, in fact, much more common than gas giants outside of the Solar System \added{(e.g., \citealt{Burke2015})}.  The radius and bulk density of Neptune (and Uranus) are often cited as context for the large number of exoplanet discoveries.  For example, exoplanets with a radius between twice the size of Earth and that of Neptune are commonly called {\it mini- or sub-Neptunes}, and planets similar in size and mass to Neptune are considered {\it Neptune-like}, since at present we cannot ascertain whether or not these distant mysterious objects share similar compositions, internal structure or evolutionary history with Neptune.  Understanding the formation, internal structure, atmosphere and evolution of Neptune is thus important, not only to investigate the physical processes at play during the formation and evolution of our own Solar System, but also for distant exoplanetary systems.  

The \ik Mission was a four-year observing campaign to search for transiting extrasolar planets, with a primary goal of determining the frequency of exoplanets with a period less than one year as a function of radius and distance from their host stars \citep{Borucki2010a}.  With almost 5 000 exoplanet candidates \citep{Coughlin2015}, over 2 000 confirmed planets (e.g., \citealt{Lissauer2014}, \citealt{Rowe2014}, \citealt{Morton2016}) and counting, it is safe to say the \ik Mission has been a resounding success (e.g., \citealt{Bedding2011}, \citealt{Chaplin2011}).   The loss of a second reaction wheel ended the \ik Mission after four years and one day of primary mission operations.  Shortly thereafter, the K2 Mission was born.  The two-wheeled K2 mission points towards fields located along the orbital plane of the \ik spacecraft to minimize torque from solar radiation pressure.  In this orientation, it achieves stable pointing using the remaining two reaction wheels supplemented with semi-periodic thruster firings \citep{Howell2014}.  The \ik Spacecraft, in a $\sim$371-day orbit around the Sun has an orbital plane similar to the Earth-Sun orbit, making some Solar System planets accessible by the K2 imager.  Of the giant planets, photometry of Jupiter and Saturn is problematic due to their apparent brightness which heavily saturates the detector, but Uranus and Neptune, while still bright enough to saturate the detector, are sufficiently faint that charge bleed from saturated pixels can be sufficiently recovered from simple aperture photometry \replaced{as a long as the charge}{as long as the charge} does not reach the edge of the detector.  Thus, high quality, photon-limited photometry of a saturated source can be obtained with the 1-meter aperture \ik instrument \citep{Gilliland2010}.

Our goal was to monitor Neptune continuously with broadband photometry offered by the \ik instrument to: study the variability of Neptune's atmosphere \citep{Simon2016}, study the Sun as a distant star through reflected light \citep{Gaulme2016} and, the subject of this paper, to search for potential oscillations that would enable seismology as a technique to probe the inner structure of the planet.  Seismology has long been considered a potentially powerful tool for probing the interiors of the giant planets \added{(e.g., \citealt{Vorontsov1976})}. Like the Sun, the fluid nature of giant planets may naturally lead to the excitation of trapped acoustic modes. The internal heat flow of Jupiter, Saturn and Neptune, respectively represent about 67, 78, and 161\% of the incident solar flux and the resulting deep convective motions have the potential to excite modes.  Detection and analysis of such oscillations is a promising technique for constraining the core mass of a giant planet, independent of the uncertainties that plague standard interior model inversion from the gravitational harmonics. The basic theory for computing giant planet oscillation frequencies and dispersion relations has been discussed as far back as \citet{Vorontsov1976} and includes work by \citet{Mosser1990} for Jupiter and \citet{Marley1991} for Saturn \added{and the more recent work of \citet{LeBihan2013} that covers a range of planetary masses and radii. }

\added{Most of the efforts dedicated to the observation of oscillations of giant planets have involved Jupiter, as it is the biggest, closest, and brightest target. There have been several attempts to detect Jovian oscillations using infrared photometry \citep{Deming1989}, Doppler spectrometry \citep{Schmider1991,Mosser1993,Mosser2000}, and careful searches for acoustic waves excited by the impact of the Shoemaker-Levy  9 comet \citep{Walter1996,Mosser1996}. In most of these campaigns, the signal-to-noise (SNR) ratio was too low or instrumental artifacts were present that inhibited any positive detection. The rapid rotation of Jupiter also limits the precision these instruments were able to obtain. While some studies have presented tantalizing upper limits (e.g, \citealt{Deming1989}) or potential detections \citep{Gaulme2011}, no definitive detection has been established.}  \deleted{Gaulme et al. 2011 reported a Doppler detection of jovian modes with peak amplitudes near frequencies of 1.2 mHz, a period of about 14 minutes. While their data window function precluded precise mode identification, they were able to measure the frequency spacing between modes, which they found to be consistent with models.} For Saturn, \citet{Marley1993} suggested that density perturbations arising from internal oscillation modes could alter ring particle orbits at resonances between orbital and mode frequencies, thereby launching waves in the rings. \citet{Hedman2013} \added{and \citet{Hedman2014}} confirmed this prediction, which suggests that at very low frequencies (periods of hundreds of minutes) Saturn is indeed oscillating \deleted{with velocity amplitudes of at least several cm/sec}. To date, modes have not been observed on Uranus or Neptune, nor have they been detected in radial velocity or broad band reflected light on any giant planet.

\replaced{The excitation of the observed modes at Jupiter and Saturn is a topic of current investigation. Leibacher1981 present an intuitive discussion of one possible excitation mechanism, while another is presented by Deming1989. Whether or not these excitation mechanisms are indeed relevant to Jupiter and Saturn--let alone Neptune--is an open question.}{\citet{Leibacher1981} discussed the possible source of excitation of modes in giant planets. Based on energetic arguments, \citet{Bercovici1987} suggested that the amplitude of Jovian modes could reach 10 to 100 cm/s. In \citet{Deming1989} demonstrated that the mechanism proposed by \cite{Goldreich1988} for solar oscillations should provide negligible amplitude for acoustic modes in the case of Jupiter, mainly because of the low vertical velocity of the convection. More recently, the different claims for p mode detection on Jupiter and particularly for f modes on Saturn have led to new investigations in this field, although no definitive results. The most promising source could be the moist convection, able to generate intermittently vertical velocities up to 100 m/s inside the storms. A possible coupling mechanism between storms and acoustic modes might exist and produced at the origin of acoustic oscillations at a detectable level on Jupiter and Saturn. Lightning is also invoked as a source of acoustic waves. Whether one of these mechanisms could be efficient on Neptune is an open question. Nevertheless, the absence of a convincing prediction for acoustic mode amplitudes on this planet doesn't preclude the possibility to detect them by observations.  The possibility of observations of Neptune with K2 presented an opportunity to try.}  

To test whether or not global modes are both excited and detectable by the ultra-high precision, high duty-cycle, integrated disc photometry offered by K2 on Neptune, we proposed to observe the planet. We present here the results of our photometric observations of the planet. \deleted{We do not attempt to motivate the plausibility of detection or the detailed mechanism by which oscillations would be observable by K2 (possibilities for the latter include the changing apparent size of the planet and albedo effects Gaulme2005). Rather} We focus on searching the relevant portion of frequency space, corresponding to periods of about 5 to 50 min, for unexpected excess power. \deleted{We defer a detailed motivation and analysis of our resulting upper limit to future work.}  \added{For a fixed albedo and radial oscillations, motion of $\sim$4 cm/sec at a period of 30 minutes would produce a change in the brightness of Neptune of $\sim$1 ppm solely from the changing apparent size of the planet.  Another detection mechanism might be feedback between atmospheric pressure perturbations and albedo, through cloud condensation effects, but such amplitude and brightness perturbation relations require further detailed analysis. Detections at ppm levels, in Fourier amplitude, have been routinely reached with \ik photometry for stars fainter than Neptune.  Our challenge was to attempt to overcome additional noise associated with the motion of Neptune across the \ik detector, which introduced significant time-correlated noise.}

In \S\ref{Observations} we present an overview of the observations and raw photometric data.  In \S\ref{method} we present our reduction methods used to extract photometric time-series and to correct for instrumental effects.  In particular, we present a novel method to disentangle the intrinsic variability of Neptune and instrumental effects produced by the high proper-motion of Neptune relative to the background star field by treating the variability of Neptune as a Gaussian process with correlated noise.  In \S\ref{results} we examine the photometric time-series to explore variability on timescales of a month to a minute with the goal of searching for and setting limits on potential oscillations intrinsic to  Neptune.

\section{Observations}\label{Observations}

The K2 C3 field provided the first opportunity to observe the planet Neptune for up to 80 days with short-cadence (1-minute) sampling\footnote{The C3 campaign had an actual duration of 69.2 days, limited by onboard data storage.}.  We were awarded sufficient pixel allocation from Guest Observer Programs GO3060 (PI: Rowe) \added{and GO3057 (PI: Gaulme)} to continuously monitor Neptune for 49 days.  K2 observed Neptune near quadrature; thus, Neptune was seen to rapidly move across the K2 field, pause and reverse course, crossing hundreds of pixels in the process.  The \ik spacecraft was designed to keep a star positioned on the same pixel for long durations.  Stable pointing reduces systematic errors in the measured brightness of a star due to effects such as intrapixel variations.  While we were able to reduce systematics in the extracted photometry the ultimate noise floor achieved was limited by the motion of Neptune across the detector.  Future observations or instrument design would benefit from maintaining the image of Neptune on a constant set of the pixels.

\begin{figure*}
\centering
\begin{tabular}{c}
\includegraphics[width=0.90\textwidth]{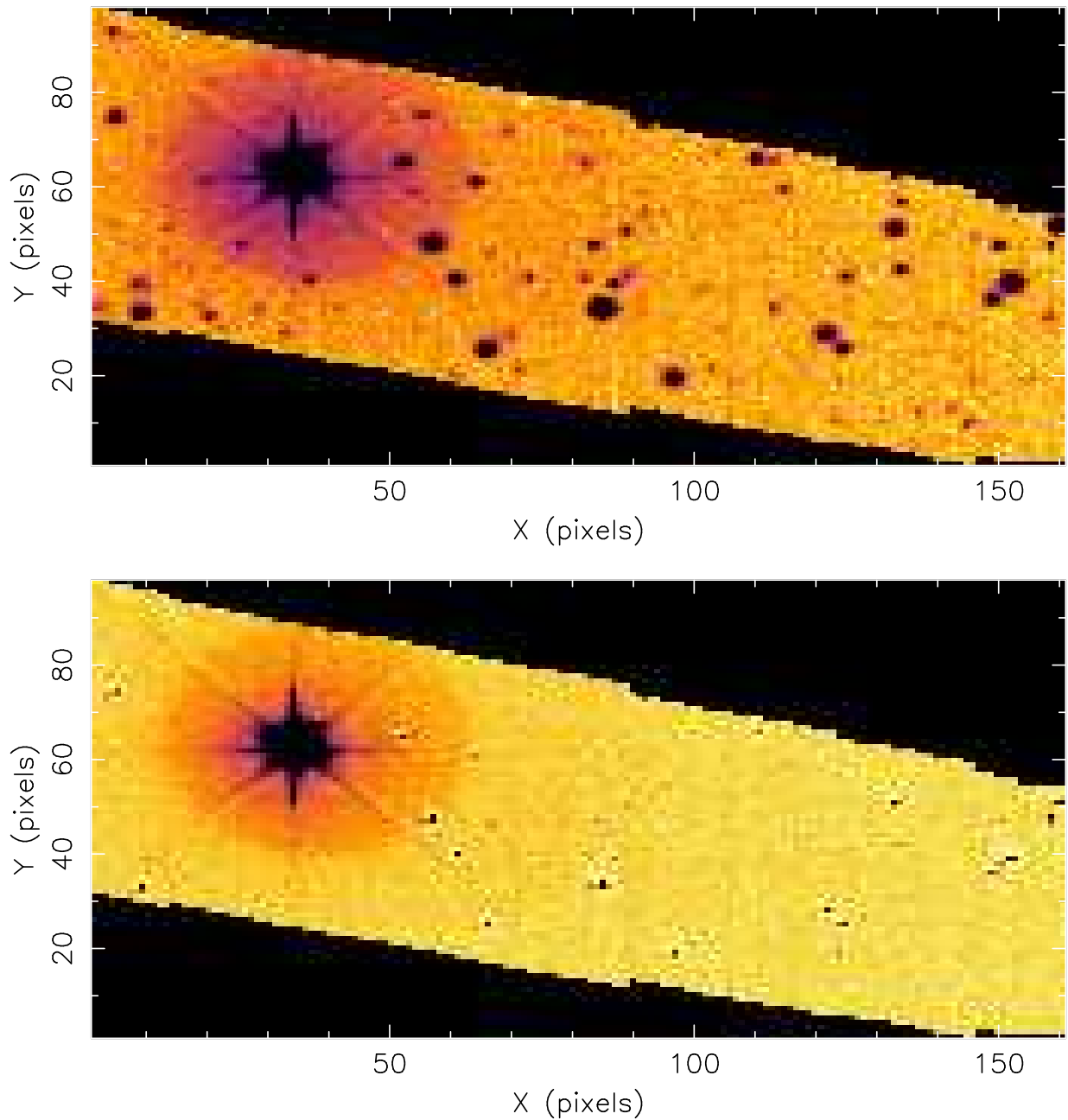}
\end{tabular}
\caption{A typical K2 short-cadence image with Neptune present is shown in the top panel.  A difference image is shown in the bottom panel.  The bright saturated target is Neptune.  The reference image was generated using images when Neptune was not captured on the subraster, thus only the background stars are removed in the difference image enabling photometry of Neptune without dilution effects from background stars.  Star residuals are due to intrapixel sensitivity variations which can be corrected for with centroid co-detrending.}
\label{fig:diffimage}
\end{figure*}

Short-cadence target pixel files were obtained from MAST\footnote{Observations from FITS files retrieved from the Mikulski Archive for Space Telescopes (MAST)}.  The Neptune short-cadence subraster was spread across 161 FITS files.  Each file contained 1 column of time-series pixel data.  Each target pixel file contains observations starting on 2014 November 15 and finishing on 2015 January 18.  We used the program {\bf kpixread}\footnote{Available on \href{https://github.com/jasonfrowe}{github.com/jasonfrowe}} \edit1{\citep[][Codebase: \url{http://doi.org/10.5281/zenodo.60297}]{jason_rowe_2016_60297} } to extract each column for each sequential observation and to produce a new FITS file containing the assembly of all pixels for a single exposure.  In total 101 579 short-cadence images were generated.  This step was necessary to use extraction and analysis processing tools such as IRAF.  The assembled images had dimensions of 162x98 pixels. An example of an assembled image is shown in the top panel of Figure \ref{fig:diffimage}.

The K2 target pixel files contain a different timestamp for each row of pixels.  The difference in time stamps is to account for different photon arrival times across the detector.  We adopted the mean time of all 161 rows as the reported time-stamp in the assembled FITS images.  The average time difference as reported in the target pixel file headers across the assembled Neptune subraster is 0.92 seconds.  Thus, reported time-stamps for Neptune in our time-series will be off by up to 0.51 seconds with the precise amount dependent on the position of Neptune.  This error is significantly smaller than the 1-minute integration time.

We scanned through the time-series of assembled short cadence images to determine when Neptune entered and left the frame.  There are $\sim$69 days of short cadence observations available but Neptune is only visible in the subraster for $\sim$49 days.  A movie\footnote{Available at \href{https://www.youtube.com/watch?v=84LDgk7l7vI}{https://www.youtube.com/watch?v=84LDgk7l7vI}} of the images was assembled that showed how the observed field changed with time.  The movie showed the star field, which revealed image jitter and several main belt asteroids that quickly move through the frame.  Neptune appears 15 days after the start of the data acquisition on 2014 December 1.  The \ik CCD detector has pixels 4x4$\arcsec$ in size, similar to the apparent size of Neptune.  The image of Neptune is heavily saturated, thus significant bleed along columns covering $\sim$20 pixels in length was observed. 

It is important to note that the charge bleed does not extend outside of the subraster or reach the edge of the detector; thus, good photometry can be extracted by aperture photometry as charge is conserved.  The motion of Neptune is observed to slow down then pause on day 41, which corresponds to when Neptune, the \ik spacecraft and the Sun were at quadrature.  After passage through quadrature, the observed motion increases until Neptune leaves the subraster on day 65.  Neptune's faint moon Nereid tracks the planet in the movie with an offset of roughly 50 pixels to the left. The bright moon Triton orbits the planet with separation of a few pixels, roughly every 5.9 days.

\section{Data Reduction Methods}\label{method} 

To extract time-series photometry of Neptune from the K2 images we performed the following tasks: i) measured the positions of field stars on each short cadence image, ii) created a reference image from images obtained during times when Neptune was not visible, iii) performed image subtraction using convolution techniques on images with Neptune present to remove field stars and iv) estimate the position of Neptune.  Photometry was measured with an aperture large enough to capture the image of Neptune, the moon Triton and column bleed from saturation.

\subsection{Centroids}

The positions of the field stars were observed to show motion that can be larger than a pixel.  Measuring the positions was important as it allowed for the rejection of frames when significant motion was detected and also allowed for the decorrelation of photometric variability due to intrapixel sensitivity changes.

To measure centroid changes we used the program {\bf allmost} \citep{Rowe2006,Rowe2008}, which was designed to extract positions and photometry from images from the MOST mission \citep{Walker2003}.  The program fits a PSF model to all stars identified in the field-of-view.  In total, 81 fields were identified and tracked.  The PSF model was a two-dimensional Gaussian.  The PSF was simultaneously  fit to all stars in the field.  The shape of the PSF was assumed to be constant across the Neptune subraster, but allowed to vary for each individual exposure.  Saturated pixels were excluded from the fits.  Valid centroids were extracted for 100 928 frames, a success rate of 99.3\%.  Some images showed extreme motion that caused the PSF fitting algorithm to fail due to significant blurring of the PSF across many pixels.

\begin{figure*}
\centering
\begin{tabular}{c}
\includegraphics[width=0.50\textwidth]{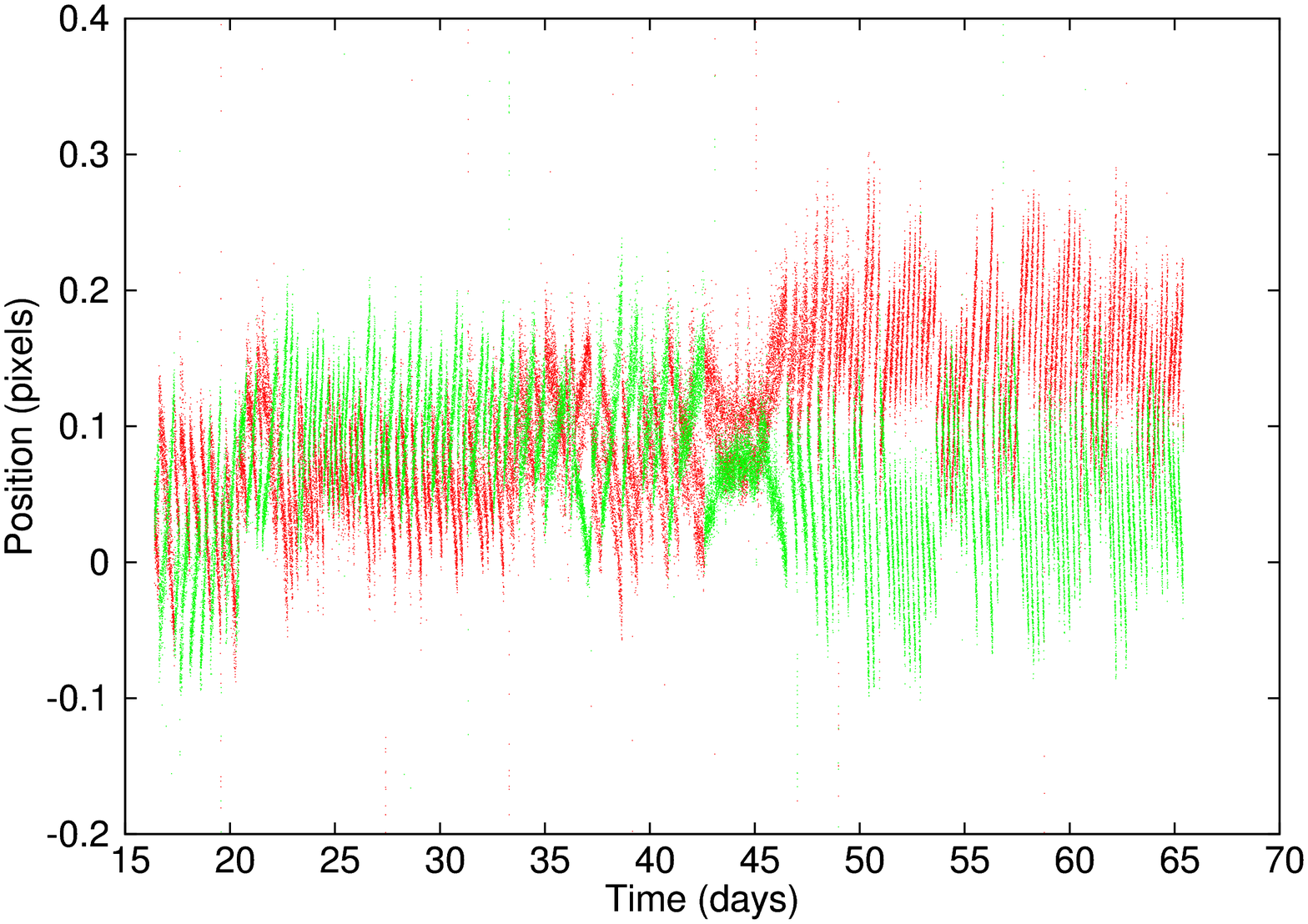}
\includegraphics[width=0.50\textwidth]{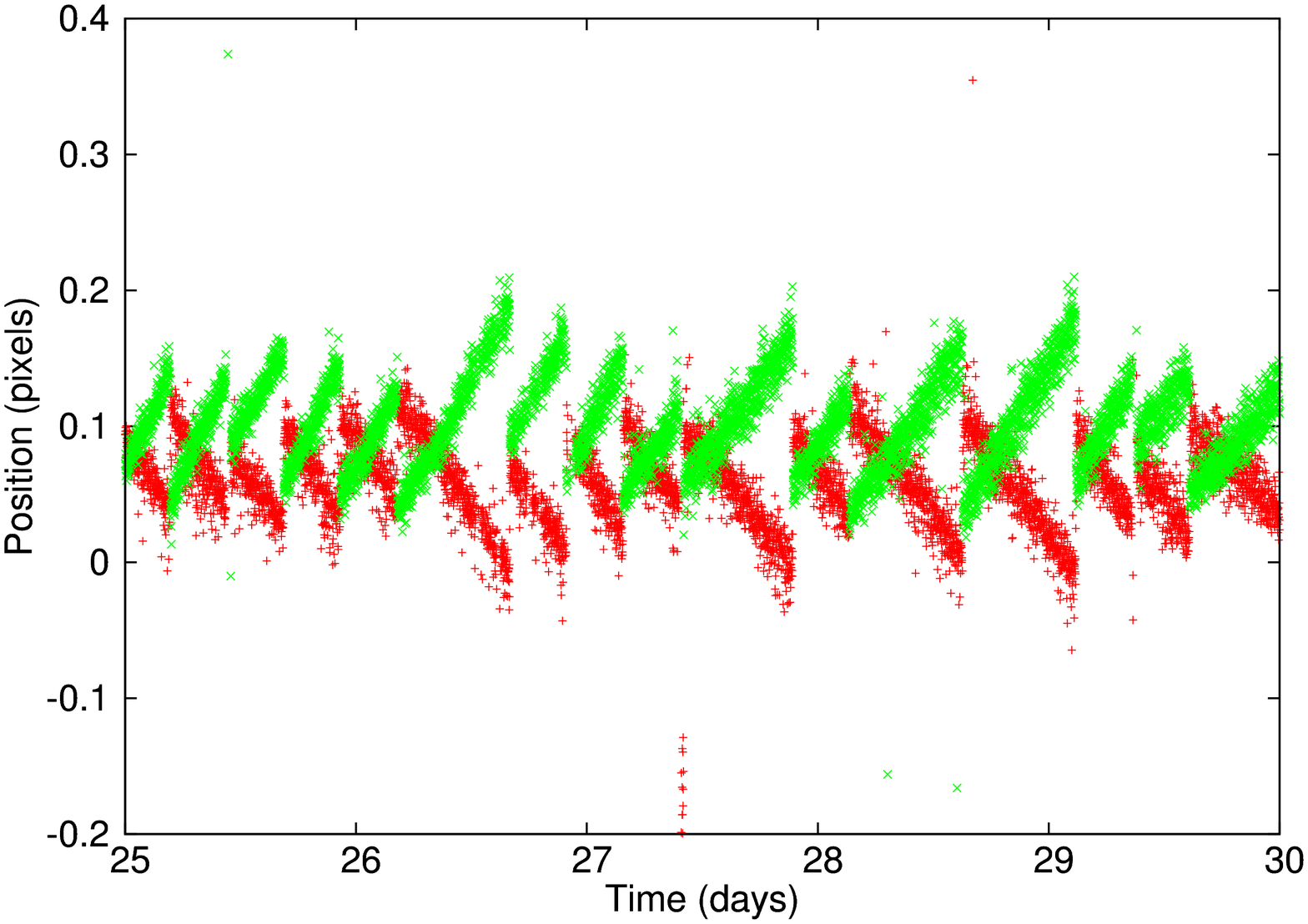}
\end{tabular}
\caption{The centroid motion of field stars measured in the Neptune SC subraster.  Larger outliers, seen every few days, are due to instrumental effects such as desaturation of reaction wheels and thruster firings, and can be corrected.  The right panel shows a small time segment of PSF centroids spanning 5 days of spacecraft motion.  Red and green markers show motion in orthogonal directions on the CCD. The abrupt changes in location are due to semi-periodic thruster firings to correct for the roll of the telescope.}
\label{fig:centroids}
\end{figure*}

The centroids for each frame were then matched to centroids from the first acquired short-cadence image (frame 1) with a model to account for shift and rotation,
\begin{equation}
\begin{array}{rl}
 x_i = S_1 + S_3 x_1 + S_5 y_1 \\
 y_i = S_2 + S_6 x_1 + S_4 y_1,
\end{array} 
\end{equation}
where $x_i$ and $y_i$ are the column and row of the assembled image.  The $x$ direction represents the primary motion of Neptune across the field of view.  The motion of frame, $i$, is relative to the first image.  The coefficients $S_1$ and $S_2$ are the shifts in position and $S_3$ through $S_6$ determine the scale and rotation of the image.  The shifts in row and column are shown in Figure \ref{fig:centroids}, with the pixel scale chosen to show the overall scatter and timescale of the motion.  The point-to-point scatter in the centroid positions demonstrates a measurement accuracy of $\sim$0.02 pixels.  The centroids show a semi-periodic behaviour of slow drift and rapid motion back to a reference position.  This behaviour is indicative of the operation of the two wheeled K2 mission: drift due to slightly unbalanced solar force until a threshold is reached, followed by thruster operation to correct the drift.

\subsection{Reference Image}

We used the first 10 000 short-cadence images (all taken prior to Neptune entering the aperture) to create a reference image to be used for image subtraction.  Images with a centroid shift greater than 0.5 pixel were excluded to minimize problems with motion smear broadening the PSF in the reference image.  The reference image was created using the software tool {\bf montage2} \citep{Stetson1987}, which handles image resampling to match image centroid changes and to create a median stacked image.

\subsection{Image Subtraction} 

We used aperture photometry to extract photometry of Neptune.  To mitigate dilution of the photometry from additional stars in the photometric aperture as Neptune traverses the subraster we used convolution techniques to match a target image to a reference image and then subtract the reference image from the target image.  The subtracted image will have the flux contributions from constant field stars removed, allowing for clean extraction of photometry of sources not present in the reference image, specifically, of Neptune.

With a reference and target image, we generated a convolution kernel \added{to match the reference image to the target image.} \replaced{that is}{The kernel is} represented as a discrete pixel array based on the method of  \citet{Bramich2008}.  One then solves for the kernel values directly using linear least squares.  The advantage of this model is that small, subpixel image drift is automatically accounted for in the kernel solution, thus the problem of interpolation for sub-pixel image registration is fitted simultaneously with changes in the PSF shape.  Pixels that are saturated or within a 15-pixel wide box centred on the position of Neptune were excluded from the kernel fit.  \added{To perform image subtraction, the reference and target image were co-alligned to the nearest integer pixel.  Then the reference image was convolved to match the target image.  The convolved image was then subtracted from the target image to produce a residual image that was used for photometry.}

The position of Neptune was determined by an average of all pixel positions with saturated counts. The column position shows a point-to-point scatter of 0.1 pixels.  The row positions show point-to-point scatter of 0.8 pixels.  The row position accuracy is degraded because of column bleed and shows some correlation with the measured flux of Neptune.

Figure \ref{fig:diffimage} shows a subtracted image at the bottom and the original image at the top.  Neptune is the bright saturated object seen to the left of each image.  The difference image shows that the stars are cleanly removed.  Some of the brightest stars show some systematic residuals in the core of the PSF that are above the expected shot noise for subtraction by factors between 1.5 and 2.  This results in a potential systematic error of 300 counts at maximum in the aperture photometry of Neptune.  The median flux measured from Neptune in a single exposure was $\sim$7.6$\times$10$^6$ e$^-$, thus, imperfect residuals from image subtraction have minimal impact on the photometry of Neptune as shot noise will dominate.

\subsection{Aperture Photometry}

We extracted Neptune photometry from difference image frame numbers 24103 through 96342, corresponding to dates 2014 December 1 through 2015 January 18.  We used an aperture box with pixel dimensions of 8$\times$30.  The width was chosen to continuously capture the flux of Triton together with Neptune to avoid the problem of Triton moving in and out of the photometric aperture during its 5-day orbit.  The height was chosen to capture column bleed across all images.  The box was centered on the pixel closest to the measured position of Neptune. Valid photometry was recovered for 71776 frames.  Frames without valid centroids or motion greater than 1 pixel were excluded from the photometric analysis.

The raw photometry is shown in Figure \ref{fig:fig3}.  There is a linear decrease in the flux that is due to the increasing distance between the \ik spacecraft and Neptune.  The 2\% variations with a period of 16 hours are due to the rotation of Neptune, with visible features appearing and disappearing from view \citep{Simon2016}.  A close-up inspection of the data in the bottom panel of Figure 3 showed a periodic trend with a timescale of $\sim$0.5 day on day 16.  This is due to intrapixel variations and is well traced by the column position of Neptune.  Thus, the timescale of this variation is not constant.  There are also sudden, abrupt changes in the flux that are correlated with column position.  We think that these jumps are due to Neptune crossing a pixel boundary and saturated bleed moving to adjacent pixels with different gain.  Sudden changes in the flux values were not seen for every crossing event.  The bottom panel of Figure \ref{fig:fig3} shows the observed behaviour we associate with pixel crossings centered on the start of day 20 with the corrections described hereafter.

\subsection{Corrections for Instrumental Effects}

We corrected for the jumps in the photometry due to pixel crossing with a piecewise line segment model.  The model, $m_i$ for each observation, can be written as, 
\begin{equation}
m_i = \left\{ \begin{array}{rl}
 a_j+b_j x &\mbox{ if $0 < x \leq1$} \\
  0 &\mbox{ otherwise}
       \end{array} \right. 
\end{equation} 
where $a_j$ and $b_j$ are fitted coefficients for the zero-point and slope for each line segment, $j$.  A segment is defined as the data points between two identified pixel crossing jump events.  The segment is defined to have a scaled length of unity, with $x$ a measure of the relative distance between the two observations.
 
Assuming Gaussian noise statistics the log-likelihood of our model matching the data is given by,
\begin{equation}
\ln p = -\frac{1}{2} r^T K^{-1} r -\frac{1}{2} \ln {\rm det} K - \frac{N}{2} \ln 2\pi
\end{equation}
where $N$ is the number of observations and $r$ defined as,
\begin{equation}
r_i = f_i - m_i,
\end{equation}
which is the difference between the photometric measurements, $f_i$, and our pixel jump model, $m_i$, for each observation index, $i$.  The transpose of $r$ is $r^T$. The co-variance matrix, $K$, was used to model the rotation modulation and long term drift observed photometrically as a Gaussian process, thus each observation was associated to a normally distributed multi-variate random variable.  Our correlated noise model is
\begin{equation}
K_{ij} = C_1 \exp \left (  \frac{-(t_i-t_j)^2}{l^2_1} \right) + C_2 \exp \left (  \frac{-(t_i-t_j)^2}{l^2_2} \right).
\end{equation}
The first component was used to model the variability of Neptune driven by rotation.  The second component models the long-term decrease in flux due to the increasing distance between the \ik spacecraft and Neptune.  The amplitude of correlated noise components is given by $C_1$ and $C_2$ for the rotation modulation and drift respectively, and the time scales are given by $l_1$ and $l_2$ respectively.

The long term drift is due to the increasing distance between Neptune and the \ik spacecraft.  The \ik spacecraft has an orbital period of 371 days compared to 60 182 days for Neptune.  Over the 49 days of short-cadence monitoring the position of Neptune can be considered fixed, whereas 
\ik will have traveled approximately 0.8 AU away from Neptune.  The distance is continuously increasing as observations from K2 were obtained near quadrature.  This change in distance corresponds to a 5\% decrease in the apparent flux, which agrees well with the K2 observations.  This allows us to set the timescale $l_2$ to the \ik orbital period.

The rotation period of Neptune is $\sim$16 hours, which corresponds to the 2\% variations observed with the same period.  We chose an initial guess of $l_1$ to be one fifth of the rotation period to model the weather variability and rotation modulation \citep{Simon2016}.  This choice of a timescale was motivated through a Fourier analysis of the times-series photometry showing the detection of harmonics up to the 5th order of the rotation period and still significantly longer than the sudden jumps that occur on 1-minute (or even shorter) timescales but short enough to capture the dynamic variability of Neptune. 

Before we fit for the pixel jump model, we identified outliers and the temporal locations of pixel-jumps.  Since not every pixel crossing event produces a jump coupled with errors in the measurement of the position of Neptune, it was difficult to precisely predict when a sudden jump occurred.  Outliers in the photometry occur due to image motion and cosmic ray hits, or may even be intrinsic to Neptune.  We did not attempt to identify the source of each outlier.  We identified outliers by comparing the change in Neptune's flux relative to the observation obtained right before or after, 
\begin{equation}
\begin{array}{rl}
 v_p = f_i - f_{i+1} \\
 v_m = f_i - f_{i-1}.
\end{array} 
\end{equation} 
If $|v_p|$ and $|v_m|$ were greater than a chosen threshold (0.0005) and $v_p/v_m > 0$ then the observation was considered an outlier and discarded from further analysis. 

To identify pixel jumps we applied a bandpass filter to the time-series photometry to remove variability with timescales longer than 0.02 day.  This timescale was chosen to remove variability in the lightcurve, but preserve the pixel jumps as excursions above and below the mean value.  To identify a jump we scanned through each time step of the detrended time-series and fit a line to the previous 10 and next 10 observations independently.  The two fitted lines give extrapolated estimates of the current measurement to compare to.  If the difference from the forward and backward time prediction is greater than 3$\sigma$ then a jump is detected and recorded.  We estimated $\sigma$ as the standard deviation of the detrended data after the application of the bandpass filter.

In total, we identified 146 jumps and 366 outliers.  We then fit our piecewise jump model and Neptune correlated noise-model to the Neptune photometric observations after the removal of outliers.  There were 296 parameters fitted: 292 parameters controlling the zero points and slopes of the line segment, two amplitudes for the noise model and two length scales to describe the variability of Neptune in the noise model.  A best-fit model was found using the L-BFGS-B code of \citet{Zhu1997}.  This code is a limited memory, quasi-Newton method that approximates the Broyden-Fletcher-Goldfarb-Shanno algorithm \added{\citep{Press1992}}.  Our best-fit model was then used to correct the pixel jumps in the data.

We corrected for intrapixel variations present in the raw time-series photometry.  Again, we treated Neptune variability as correlated noise but used a longer timescale of 0.17 day to avoid degeneracy between the noise model and the timescale of pixel crossings which varies from stationary to $\sim$10 pixels/day.

Our intrapixel model describes the photometric flux vs pixel location in row and column.  We indexed the current row and column position as a function of time.  Thus, if Neptune returns to the same pixel later in time, that pixel event is given a unique index.  This allows the use of a linear model to describe the intrapixel variations and pixel-to-pixel gain variations.  Our model, $pm$, is a series of sinusoids -- one for each pixel index, $j$, which can be written as,
\begin{equation}
pm_j = A_j \sin (2\pi x + \phi_j)
\end{equation}
where $j$ is the pixel index, $A_j$ and $\phi_j$ are the amplitude and phase of the correction and $x$ is the relative column position.  To avoid discontinuities in the model, we linearly interpolated the applied amplitude between adjacent pixels to create a smooth function.  We use the same L-BFGS-B algorithm as above to find a best-fit intrapixel model.

\begin{figure*}
\centering
\begin{tabular}{c}
\includegraphics[width=0.90\textwidth]{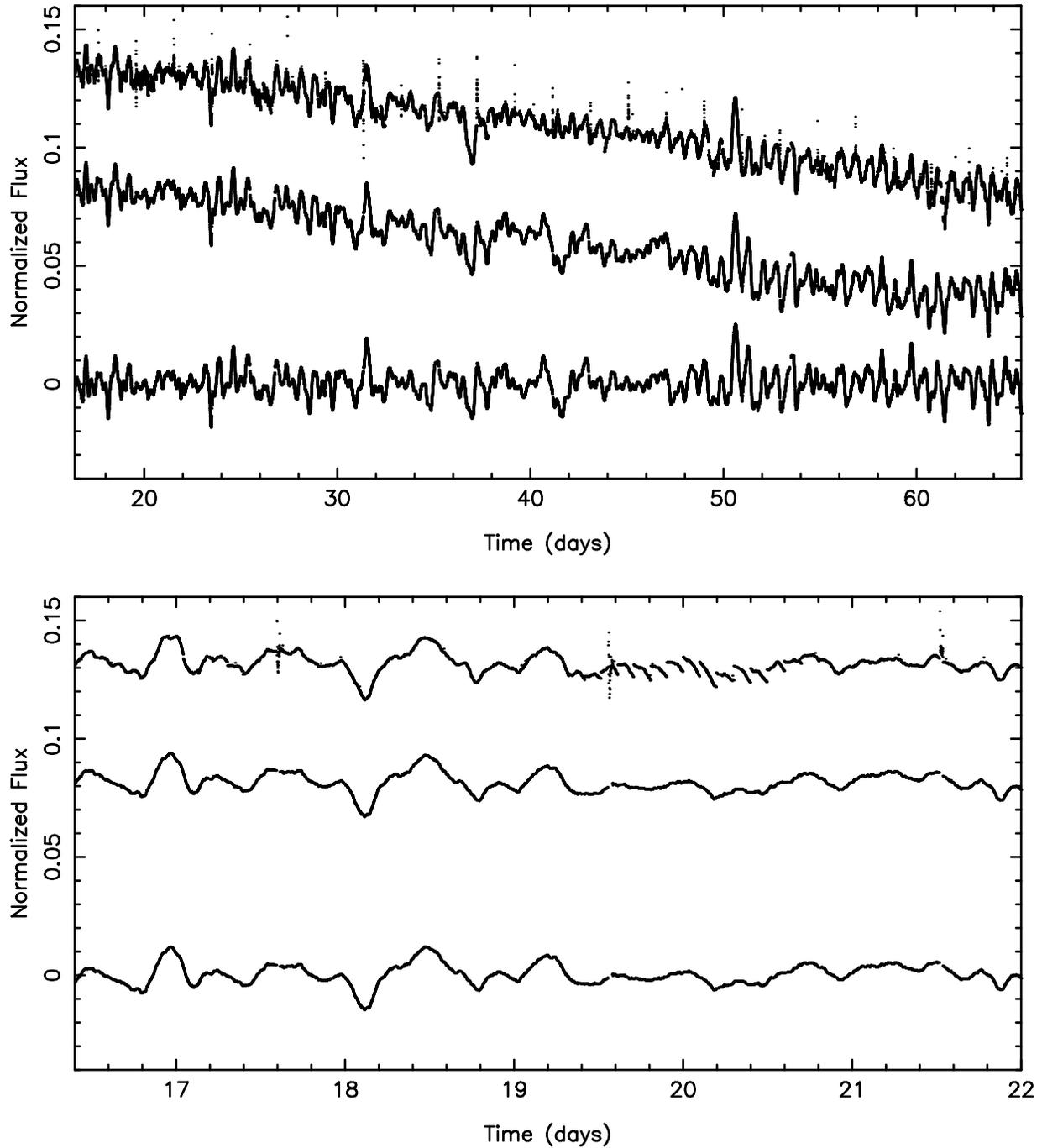}
\end{tabular}
\caption{The top panel shows the raw, corrected and distance corrected photometry.  The raw and corrected photometry have been arbitrarily offset by 0.11 and 0.06 in normalized flux for plotting purposes.  The time is relative to the start time of short-cadence acquisition of the Neptune subraster, which corresponds to BJD=2456977.09331.  The bottom panel shows the same photometry as the top panel but from just after day 16 to day 22, representing the first $\sim$6 days of short-cadence photometry.  The zoom showcases the corrections of the photometric variations due to intrapixel variations as seen during day 16 and pixel-jumps as seen near day 20.}
\label{fig:fig3}
\end{figure*}

The top panel of Figure \ref{fig:fig3} shows the corrected light-curve for Neptune.  The bottom panel of Figure \ref{fig:fig3} shows a comparison of the raw and corrected light-curves for a small segment of the data.  The pixel jumps and intrapixel variations are well corrected.  Table \ref{tab:phot} contains our distance-corrected adopted time-series observations for analysis as plotted in Figure \ref{fig:fig3}.

\section{Results}\label{results}

We removed the long-term trend in the corrected photometry due to the distance effect.  Our adopted time-series photometry used for the analysis is presented as the bottom lightcurve in Figure \ref{fig:fig3}.  In Figure \ref{fig:fft} we show the power density and windowed Fourier transform of the photometric time-series.  The Fourier transform was computed using an FFT.  The time-series was resampled onto an equally spaced grid using linear interpolation, which included accounting for gaps in the time-series.  The data was zero-padded to achieve an oversampling factor of 5.  The power density spectrum (PDS) is presented in the bottom panel of Figure \ref{fig:fft} and reveals a few features of the variability observed from the Neptune observations: $1/\nu$ noise, rotation modulation and solar variability.  There is a clear decrease in PDS with increasing frequency.  The red line in Figure \ref{fig:fft} is an estimate of the mean PDS (see \S\ref{psearch} below) which shows a change in slope around 60 \uhz.  This change is likely related to observation of two sources of variability: one intrinsic to the instrument and the other related to astrophysical processes such as solar granulation and Neptune weather.  The $1/\nu$ behaviour observed for frequencies larger than $\sim$60 \uhz\ is likely instrumental noise related to the motion of Neptune across the subraster and the motion of incident flux across hundreds of different pixels.  The windowed Fourier transform shows a time-frequency representation of the data and is shown in the top panel of Figure \ref{fig:fft}.  The figure was created using a running tapered window with a length of 5 days.  For frequencies above $\sim$60 \uhz\ the amplitude of the noise floor is variable with time and strongly correlated with the velocity of Neptune across the detector.  The lowest levels are observed near day 41, which corresponds to when the motion of Neptune is minimized.  The additional noise from motion has a $1/\nu$ dependence and ultimately sets the detection limit for frequencies shorter than $\sim$60 \uhz. 

The excess power between 10 and 20 \uhz\ is the rotation modulation from Neptune (See Figure 2 from \citep{Simon2016}).  Localized bright clouds high in the atmosphere reflect light back towards the observer before the onset of scattering and absorption due to haze and methane deeper in the atmosphere.  The clouds generally trace the zonal wind velocity and rotation period \citep{Simon2016}.  The observed flux is modulated as the clouds evolve or rotate from view.  The first through fifth harmonics of the rotation period were detected, showing the non-sinusoid shape of the rotation signature.  Near 3000 \uhz\ there are a handful of significant frequencies detected and an excess of power centred on these frequencies is detected.  This variability is due to solar p mode oscillations seen in reflected light.  The detection of solar oscillations allows for the study of the Sun as a distant star as the observed signal represents an integrated disk measurement \citep{Gaulme2016}.  There are two peaks at very high frequency beyond 7000 \uhz.  These are instrumental features that were first recognized in short-cadence observations from the 4-year \ik mission and are thought to be related to the inverse of the long-cadence ($\sim$30 minute) sample time.  The location of these frequencies measured from K2 observations has changed relative to the \ik mission, suggesting evolution of the instrument and the root cause of this signal.

The primary goal of this paper was to search for oscillations from Neptune.  The idea is that, as observed in the Sun, convection-driven p modes can have sufficient amplitude and coherence that such behaviour could be detected in disk-integrated light from Neptune.  \added{While p-mode excitation in astrophysical fluids, such as present in the Sun, is qualitatively understood (e.g., \citealt{Goldreich1988}) it should be noted that convection in gas-giants may be too slow to drive oscillations \citep{Deming1989}.}  \replaced{A simple}{However, any} detection of oscillation frequencies or excess power would enable the study of the interior of the planet analogous to studies of the Sun's interior via helioseismology.  The detection or the placing of a significant upper limit is also important for the planning of future instruments or observations to observe the global oscillations.  The existence of resonant modes of oscillations in a planet requires a trapping mechanism for the waves. As first demonstrated by \citet{Vorontsov1976}, the atmospheres of the giant planets indeed reflect acoustic waves, but only if their frequency is below a cut-off value. Detailed calculations of the vertical propagation of acoustic waves by  \citet{Mosser1995} show that the level at which the waves are reflected is a strong function of their frequency. In all four giant planets, it turns out that most waves are reflected just below the tropause and above the main cloud deck. In the case of Neptune, the maximum frequency for resonant acoustic modes is about 3000 \uhz\ \citep[see][Fig. 2]{Gaulme2015}.  The possible acoustic waves excited at larger frequencies are not trapped by Neptune's troposphere and get dissipated in the stratosphere. More precisely, after increasing monotonically in the upper troposphere, the cut-off frequency reaches a maximum at the tropopause, and decreases to a plateau at about 2000 \uhz\ in the mid-stratosphere. As analyzed by \citet{Mosser1995} for Jupiter, it means that waves with frequencies between 2000 and 3000 \uhz\ can leak into the troposphere by tunneling effect, making their trapping less efficient. Regarding the lower limit, the acoustic cut-off profile shows that waves with frequencies less that 800 \uhz\ are trapped deeper than pressure levels of about 10 bars. Even though radiative transfer in these planets is not fully constrained, it is very likely that optical observations do not probe that deep. Waves at frequencies lower than 800 \uhz\ are thus evanescent at altitudes probed by optical observations. It it therefore reasonable to expect Neptune oscillations' maximum amplitude to be between 800 and 2000 \uhz.

Regarding the mean frequency spacing between mode overtones, it is expected to range from 198 to 213 $\mu$Hz according to internal structure models \citep[][]{Gaulme2015}. In addition, because of Neptune's rapid rotation, acoustic modes of  non-radial oscillations with azimuthal order $m$ not equal to zero split apart into  $+m$ and $-m$ peaks separated by $2m$ times the inverse rotation period, i.e. $\approx 2m\times 17.4\ \mu$Hz. 

\subsection{Search for Excess Power}

The power spectral density (Fig. \ref{fig:fft}) does not display any excess power typical of global oscillations in the [800, 2000] \uhz\ range, and nowhere else except for the solar oscillations. We also searched for Neptune's oscillations in the envelope of the autocorrelation (EACF) of the time-series, filtered in the expected frequency domain \citep{Roxburgh2006, Mosser2009b}. This approach allows for deriving the mean large separation of a solar-like oscillation spectrum in a blind way without prior information. It has shown to be efficient in cases of low SNR \citep[e.g.][]{Mosser2009c, Gaulme2010, Mosser2010}. The reliability of the result is given by an $H_0$ test: when the EACF is above a threshold level, the null hypothesis can be rejected, implying that a signal might have been detected. 

The envelope of the autocorrelation displays a maximum in between 1.96 and 2.15 hours in the frequency range $[600, 1400]\ \mu$Hz (Fig. \ref{fig:eacf}). From the $H_0$ test, the likelihood for this peak to be a signal is about 95\,\%. This would correspond with a large frequency spacing ranging from 258 to 283 \uhz\ if it is the result of Neptune's oscillations. However, if this excess of power in the autocorrelation diagram is generated by Neptune's oscillations, we should find peaks corresponding to its rotation period (16-17h) and some harmonics, as modes are split by it. Figure \ref{fig:eacf} does not exhibit significant maxima at 16 or 8 hours. As a comparison, we displayed the correlation diagram up to frequencies including the solar oscillations. We clearly detect the Sun's large separation, as well as secondary peaks due to the various overtones and the separation between $\ell=0$ and $\ell=1$ modes. Finally, we investigated whether a structure could be detected in the \'echelle diagram corresponding with a $\approx 280\ \mu$Hz and nothing was observed, making it useless to reproduce in the paper. The signal in the $[600, 1400]\ \mu$Hz range does not show similar features, and we conclude this 2-hour signal it is likely spurious and likely related to incomplete corrections of instrumental signals due to the motion of Neptune.  

\subsection{Search for Significant Frequencies}\label{psearch}

The PDS was observed to significantly drop off the 40-80 \uhz\ region that is dominated by power leakage and variability associated with weather and rotation of Neptune, thus we conducted our search for significant frequencies and/or excess power on timescales shorter than $10^4$\,s (frequencies above 100 \uhz). The search for individual frequencies assumes that any such signal is coherent over a few days (as indicated by stellar oscillations, e.g. \citealt{Appourchaux2008}).  The search for excess power assumes that an envelope of power would be produced that is similar to the observation of low-power p modes observed in the Sun.

To estimate the significance of any frequency from the Fourier analysis we fit a semi-Lorentzian model to the observed PDS of the Neptune photometric time-series.  The addition of a Gaussian centred on the observed location of the solar p modes was also used to model excess power in that range.  The adopted model was,
\begin{equation}
P(f) = B + \sum_i^N \frac{A_i}{1-(\theta_i f)^{\alpha_i}} + D\ {\rm exp}\left( \frac{-(f_{\rm max}-f)^2}{2\sigma^2} \right), 
\end{equation}
where $B$ is the noise floor (ppm$^2$/\uhz), $A_i$ is the amplitude of the semi-Lorentzian, $\theta_i$ is the characteristic timescale and $\alpha_i$ is the decay.  The Gaussian component has an amplitude, $D$, centred on frequency, $f_{max}$ with width, $\sigma$.  We found that $N=2$ components provided a reasonable fit.  Best fit parameters were obtained with a Levenberg-Marquardt chi-square minimization routine \citep{More1980}.  Our best fit parameters are listed in Table \ref{tab:fitpars}.

Figure \ref{fig:fftnorm} shows the amplitude spectrum after normalization with our power-density model.  We observed that a histogram of normalized power shows a log-linear trend, this means that Gaussian chi-square statistics can be used to determine the probability of the power associated with any frequency may be due to noise \citep{Gabriel2002}.   Equation 12 from \citet{Gabriel2002} was used to estimate the probability of a frequency with power $m$ times the mean,
\begin{equation}
P(m) = 1 - (1-e^{-m})^{\rho {\rm N}},
\end{equation}
with, N=52429, the number of Nyquist sampled frequencies, $\rho = 2.6$ to account for 5 times oversampling. We solved to find $m$ with $P(m)=1.9\times10^{-5}$, which corresponds to a false-alarm detection of no more than one frequency of significance.   We find $m=9.89$, which is plotted in Figure \ref{fig:fftnorm}.  There is a clear detection of individual solar p modes around 3000 \uhz\ and the \ik instrumental frequencies above 7000 \uhz.  No other frequencies above 100 \uhz\ are detected.  We found no candidate frequencies that could be due to intrinsic oscillations of Neptune.  Using our value for $m$ we can place an upper limit of $\sim$5 ppm at 1000 \uhz\ for the detection of a coherent signal\footnote{divide by $\sqrt{9.89}$ to get the mean level}.  Our noise limit ranges from $\sim$100 ppm at 100 \uhz\ to $\sim$2.3 ppm at 3000 \uhz.  The red line in Figure \ref{fig:fftnorm} is a running mean with a bin width of 30 \uhz.  There is no strong evidence of excess power above 100 \uhz.  Thus, we conclude that we do not find any evidence of oscillations in broadband photometry of Neptune obtained by the K2 mission.

\added{For fixed albedo a brightness change of 5 ppm is equivalent to a change in the radius of Neptune of 62 m.  On a 17 minute timescale ($\sim$1000 \uhz) such a radius change translates to a radial velocity semi-amplitude of 38 cm/s.  Our intended goal was to probe variability at amplitudes lower than 1 ppm which would set radial upper limits of better 8 cm/s.}

%Our detected frequencies are listed in Table 3. 
% 2899.2, 2963.3, 3232.8, 3303.7, 3915.4 uHz - like Solar p mode detections.

%\begin{itemize}
%\item FT means
%\item excess power
%\item excess peaks in power spectrum 
%\item signatures of Triton and other moons in the data
%\end{itemize}

\section{Summary and Conclusions}

We have presented 49 days of continuous broadband photometry of the planet Neptune from the K2 mission.  The photometry shows 2\% variability associated with the presence of evolving bright clouds and the rotation of Neptune.  There is also a clear detection of solar p mode oscillations in reflected light which demonstrates the overall data quality and the ability to detect oscillations with amplitudes of $\sim$ppm levels in the Fourier domain \citep{Gaulme2016}.  Our goal was to detect oscillations intrinsic to Neptune, but our search was unsuccessful.  The noise floor, while impressively low, was ultimately set by the motion of Neptune across the field of view crossing hundreds of pixels coupled with pointing jitter and pixel-to-pixel and intrapixel gain variations.

We were able to correct instrumental artifacts present in the raw K2 aperture photometry for a moving object by using difference imaging to mitigate dilution of the photometric signal from background stars passing through a photometric aperture centered on the observed position of Neptune.  Additionally, by treating the variability of Neptune as correlated Gaussian noise we were able to model intrapixel variations and sudden changes in the photometry that were correlated with the motion of Neptune.

\begin{figure*}
\centering
\begin{tabular}{c}
\includegraphics[width=0.90\textwidth]{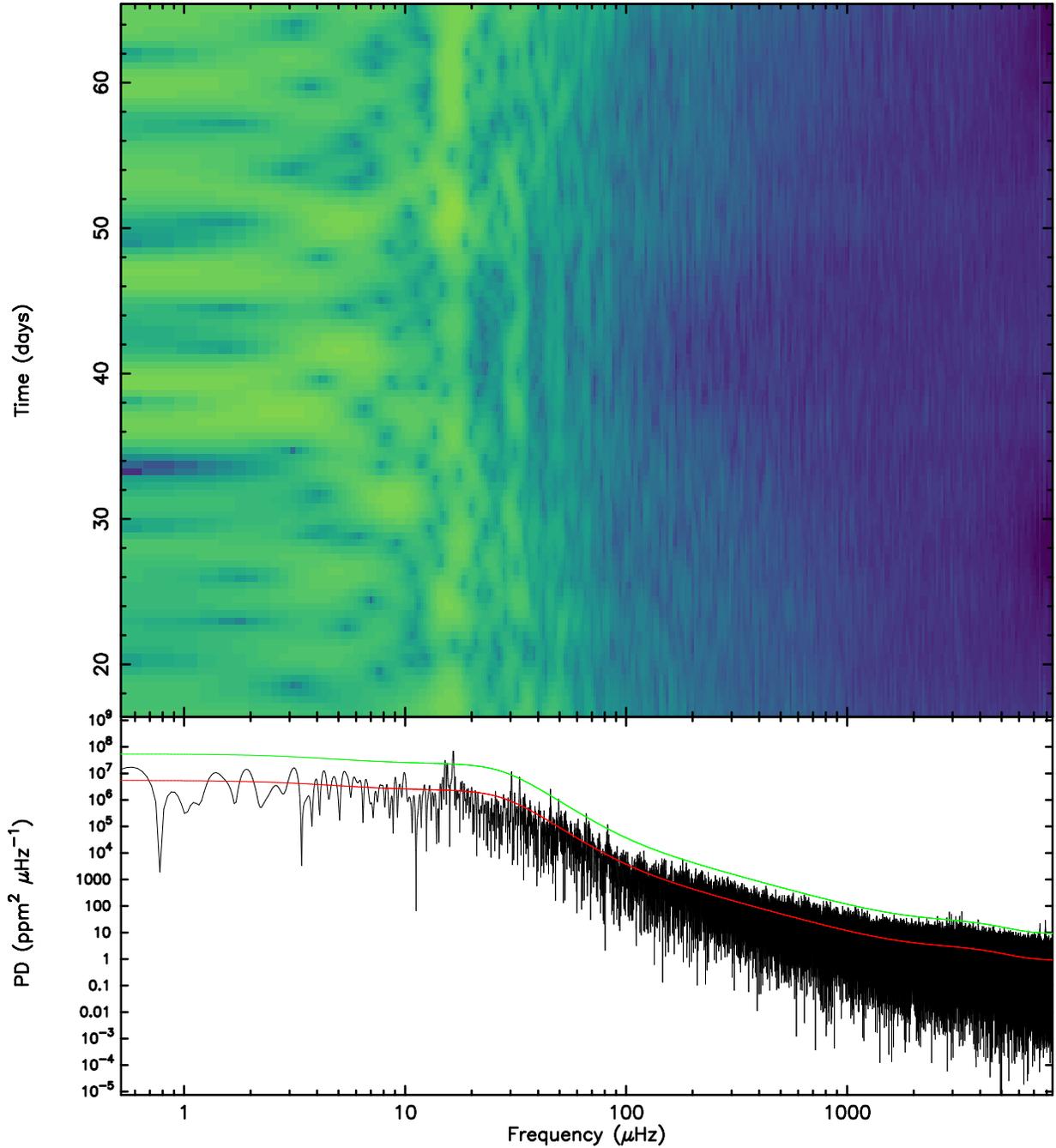}
\end{tabular}
\caption{The bottom panel plots the power spectral density (PSD) of the distance corrected photometry.  The calculation  is described in \S4. The red line shows the fitted PSD model based on Equation 8 which is an estimate of the mean.  The green line is 9.89 times the PSD model and presents our adopted cut to identify significant frequencies as described in \S4.1.  The top panel shows a windowed PSD to demonstrate variability in PSD as a function of time.  Light, green colour represent larger values of the PSD and darker bluer colours correspond to lower values of the PSD.
}
\label{fig:fft}
\end{figure*}

\begin{figure*}
\centering
\begin{tabular}{c}
\includegraphics[width=0.90\textwidth]{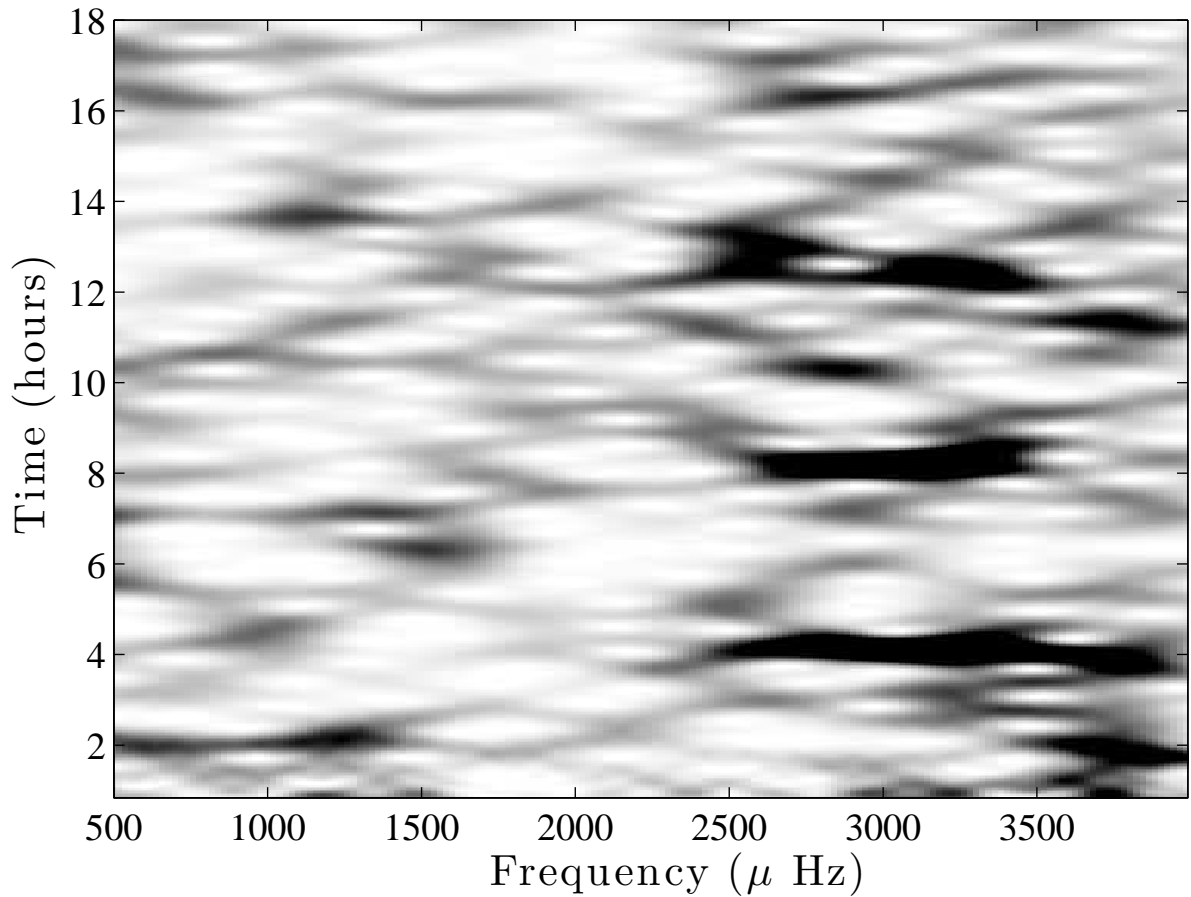}
\end{tabular}
\caption{Envelope of the autocorrelation function (EACF) as function of frequency ($x$-axis) and time ($y$-axis). The EACF was computed every 20~$\mu$Hz from 500 to 4000~$\mu$Hz, by filtering the time-series in a frequency bandpass of 500~$\mu$Hz. The darker a region is, the larger the correlation. Solar modes are clearly visible in between 2500 and 3500~$\mu$Hz at approximately 4, 8 and 13 hours.
}
\label{fig:eacf}
\end{figure*}

\begin{figure*}
\centering
\begin{tabular}{c}
\includegraphics[width=0.90\textwidth]{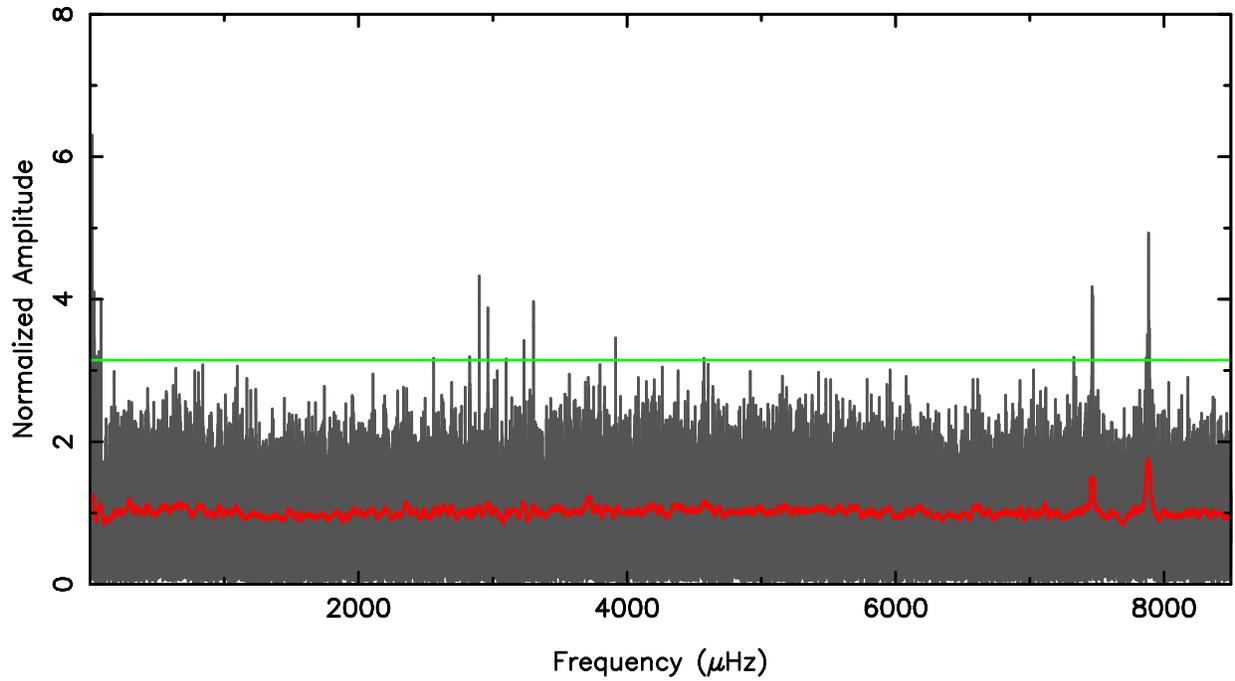}
\end{tabular}
\caption{Normalized amplitude spectrum of Neptune photometry based on our power density spectrum model which includes fitting for excess power due to Solar p modes.  The green line marks our confidence level to have a false-alarm detection of less than one.  The red line is a running average with a binwidth of 30 \uhz. Significant frequencies are seen a low frequencies (less than 100 \uhz) due to weather and rotation of Neptune, Solar modes near 3000 \uhz and high-frequency instrumental signals between 7000 and 8000 \uhz.   
}
\label{fig:fftnorm}
\end{figure*}

While our search for intrinsic convection-driven pulsations in Neptune was unsuccessful we have learned valuable lessons that can be applied to future photometric observations Solar System objects.  In 2016, K2 observed the planet Uranus, which provides a rare opportunity to obtain a long un-interrupted photometric time-series of the planet to repeat our Neptune experiment: to characterize the weather and cloud timescales, to study the Sun as a distant star through reflected light and to search for intrinsic Uranus oscillations.  Our experience with Neptune photometry has also taught us how to reduce the impact of instrumental effects from moving objects observed by K2 and how to plan future missions or photometric campaigns.  The obvious conclusion is that it is beneficial to keep the position of Neptune, or any target, located on the same pixel.  The windowed FT in Figure \ref{fig:fft} shows a clear decrease in the high frequency noise floor around day 25.  This corresponds to when K2 and Neptune passed through quadrature and the planet motion across the detector was minimized.  \added{A \ikt-like instrument capable of tracking Neptune would be capable of probing the intrinsic variability of Neptune at ppm levels on timescales of 30 minutes and shorter.}  Our detection limits in the Fourier domain with 49 days of broadband photometry sets a benchmark for future experiments to search for oscillation patterns with integrated disk photometry.   

\acknowledgments

%J.F.R. would like to thank Knicole Colon, Douglas Caldwell and Michael Haas for useful discussions regarding data artifacts present in Kepler short-cadence observations.

J.F.R. acknowledges NASA grant NNX14AB82G issued through the Kepler Participating Scientist Program. Funding for the Stellar Astrophysics Centre is provided by The Danish National Research Foundation (Grant agreement No. DNRF106). The research is supported by the ASTERISK project (ASTERoseismic Investigations with SONG and {\it Kepler}) funded by the European Research Council (Grant agreement No. 267864). VSA. acknowledges support from VILLUM FONDEN (research grant 10118).

\added{ \software{allmost (Rowe et al. 2006, 2008), montage2 (Stetson 1987), L-BFGS-B code (Zhu et al. 1997)}
}

\bibliography{AstroRefs.bib}

\begin{thebibliography}{}
\expandafter\ifx\csname natexlab\endcsname\relax\def\natexlab#1{#1}\fi

\bibitem[{{Appourchaux} {et~al.}(2008){Appourchaux}, {Michel}, {Auvergne},
  {Baglin}, {Toutain}, {Baudin}, {Benomar}, {Chaplin}, {Deheuvels}, {Samadi},
  {Verner}, {Boumier}, {Garc{\'{\i}}a}, {Mosser}, {Hulot}, {Ballot}, {Barban},
  {Elsworth}, {Jim{\'e}nez-Reyes}, {Kjeldsen}, {R{\'e}gulo}, \&
  {Roxburgh}}]{Appourchaux2008}
{Appourchaux}, T., {Michel}, E., {Auvergne}, M., {et~al.} 2008, \aap, 488, 705

\bibitem[{{Bedding} {et~al.}(2011){Bedding}, {Mosser}, {Huber},
  {Montalb{\'a}n}, {Beck}, {Christensen-Dalsgaard}, {Elsworth},
  {Garc{\'{\i}}a}, {Miglio}, {Stello}, {White}, {De Ridder}, {Hekker}, {Aerts},
  {Barban}, {Belkacem}, {Broomhall}, {Brown}, {Buzasi}, {Carrier}, {Chaplin},
  {di Mauro}, {Dupret}, {Frandsen}, {Gilliland}, {Goupil}, {Jenkins},
  {Kallinger}, {Kawaler}, {Kjeldsen}, {Mathur}, {Noels}, {Silva Aguirre}, \&
  {Ventura}}]{Bedding2011}
{Bedding}, T.~R., {Mosser}, B., {Huber}, D., {et~al.} 2011, \nat, 471, 608

\bibitem[{{Bercovici} \& {Schubert}(1987)}]{Bercovici1987}
{Bercovici}, D., \& {Schubert}, G. 1987, \icarus, 69, 557

\bibitem[{{Borucki} {et~al.}(2010){Borucki}, {Koch}, {Basri}, {Batalha},
  {Brown}, {Caldwell}, {Caldwell}, {Christensen-Dalsgaard}, {Cochran},
  {DeVore}, {Dunham}, {Dupree}, {Gautier}, {Geary}, {Gilliland}, {Gould},
  {Howell}, {Jenkins}, {Kondo}, {Latham}, {Marcy}, {Meibom}, {Kjeldsen},
  {Lissauer}, {Monet}, {Morrison}, {Sasselov}, {Tarter}, {Boss}, {Brownlee},
  {Owen}, {Buzasi}, {Charbonneau}, {Doyle}, {Fortney}, {Ford}, {Holman},
  {Seager}, {Steffen}, {Welsh}, {Rowe}, {Anderson}, {Buchhave}, {Ciardi},
  {Walkowicz}, {Sherry}, {Horch}, {Isaacson}, {Everett}, {Fischer}, {Torres},
  {Johnson}, {Endl}, {MacQueen}, {Bryson}, {Dotson}, {Haas}, {Kolodziejczak},
  {Van Cleve}, {Chandrasekaran}, {Twicken}, {Quintana}, {Clarke}, {Allen},
  {Li}, {Wu}, {Tenenbaum}, {Verner}, {Bruhweiler}, {Barnes}, \&
  {Prsa}}]{Borucki2010a}
{Borucki}, W.~J., {Koch}, D., {Basri}, G., {et~al.} 2010, Science, 327, 977

\bibitem[{{Bramich}(2008)}]{Bramich2008}
{Bramich}, D.~M. 2008, \mnras, 386, L77

\bibitem[{{Burke} {et~al.}(2015){Burke}, {Christiansen}, {Mullally}, {Seader},
  {Huber}, {Rowe}, {Coughlin}, {Thompson}, {Catanzarite}, {Clarke}, {Morton},
  {Caldwell}, {Bryson}, {Haas}, {Batalha}, {Jenkins}, {Tenenbaum}, {Twicken},
  {Li}, {Quintana}, {Barclay}, {Henze}, {Borucki}, {Howell}, \&
  {Still}}]{Burke2015}
{Burke}, C.~J., {Christiansen}, J.~L., {Mullally}, F., {et~al.} 2015, \apj,
  809, 8

\bibitem[{{Chaplin} {et~al.}(2011){Chaplin}, {Kjeldsen},
  {Christensen-Dalsgaard}, {Basu}, {Miglio}, {Appourchaux}, {Bedding},
  {Elsworth}, {Garc{\'{\i}}a}, {Gilliland}, {Girardi}, {Houdek}, {Karoff},
  {Kawaler}, {Metcalfe}, {Molenda-{\.Z}akowicz}, {Monteiro}, {Thompson},
  {Verner}, {Ballot}, {Bonanno}, {Brand{\~a}o}, {Broomhall}, {Bruntt},
  {Campante}, {Corsaro}, {Creevey}, {Do{\u g}an}, {Esch}, {Gai}, {Gaulme},
  {Hale}, {Handberg}, {Hekker}, {Huber}, {Jim{\'e}nez}, {Mathur}, {Mazumdar},
  {Mosser}, {New}, {Pinsonneault}, {Pricopi}, {Quirion}, {R{\'e}gulo},
  {Salabert}, {Serenelli}, {Silva Aguirre}, {Sousa}, {Stello}, {Stevens},
  {Suran}, {Uytterhoeven}, {White}, {Borucki}, {Brown}, {Jenkins}, {Kinemuchi},
  {Van Cleve}, \& {Klaus}}]{Chaplin2011}
{Chaplin}, W.~J., {Kjeldsen}, H., {Christensen-Dalsgaard}, J., {et~al.} 2011,
  Science, 332, 213

\bibitem[{{Coughlin} {et~al.}(2015){Coughlin}, {Mullally}, {Thompson}, {Rowe},
  {Burke}, {Latham}, {Batalha}, {Ofir}, {Quarles}, {Henze}, {Wolfgang},
  {Caldwell}, {Bryson}, {Shporer}, {Catanzarite}, {Akeson}, {Barclay},
  {Borucki}, {Boyajian}, {Campbell}, {Christiansen}, {Girouard}, {Haas},
  {Howell}, {Huber}, {Jenkins}, {Li}, {Patil-Sabale}, {Quintana}, {Ramirez},
  {Seader}, {Smith}, {Tenenbaum}, {Twicken}, \& {Zamudio}}]{Coughlin2015}
{Coughlin}, J.~L., {Mullally}, F., {Thompson}, S.~E., {et~al.} 2015, ArXiv
  e-prints, arXiv:1512.06149

\bibitem[{{Deming} {et~al.}(1989){Deming}, {Mumma}, {Espenak}, {Jennings},
  {Kostiuk}, {Wiedemann}, {Loewenstein}, \& {Piscitelli}}]{Deming1989}
{Deming}, D., {Mumma}, M.~J., {Espenak}, F., {et~al.} 1989, \apj, 343, 456

\bibitem[{{Gabriel} {et~al.}(2002){Gabriel}, {Baudin}, {Boumier},
  {Garc{\'{\i}}a}, {Turck-Chi{\`e}ze}, {Appourchaux}, {Bertello}, {Berthomieu},
  {Charra}, {Gough}, {Pall{\'e}}, {Provost}, {Renaud}, {Robillot}, {Roca
  Cort{\'e}s}, {Thiery}, \& {Ulrich}}]{Gabriel2002}
{Gabriel}, A.~H., {Baudin}, F., {Boumier}, P., {et~al.} 2002, \aap, 390, 1119

\bibitem[{{Gaulme} {et~al.}(2015){Gaulme}, {Mosser}, {Schmider}, \&
  {Guillot}}]{Gaulme2015}
{Gaulme}, P., {Mosser}, B., {Schmider}, F.-X., \& {Guillot}, T. 2015,
  {Seismology of Giant Planets}, ed. V.~C.~H. {Tong} \& R.~A. {Garc{\'{\i}}a},
  189--202

\bibitem[{{Gaulme} {et~al.}(2011){Gaulme}, {Schmider}, {Gay}, {Guillot}, \&
  {Jacob}}]{Gaulme2011}
{Gaulme}, P., {Schmider}, F.-X., {Gay}, J., {Guillot}, T., \& {Jacob}, C. 2011,
  \aap, 531, A104

\bibitem[{{Gaulme} {et~al.}(2010){Gaulme}, {Deheuvels}, {Weiss}, {Mosser},
  {Moutou}, {Bruntt}, {Donati}, {Vannier}, {Guillot}, {Appourchaux}, {Michel},
  {Auvergne}, {Samadi}, {Baudin}, {Catala}, \& {Baglin}}]{Gaulme2010}
{Gaulme}, P., {Deheuvels}, S., {Weiss}, W.~W., {et~al.} 2010, \aap, 524, A47+

\bibitem[{{Gaulme et al.}(2016)}]{Gaulme2016}
{Gaulme et al.} 2016, in preparation

\bibitem[{{Gilliland} {et~al.}(2010){Gilliland}, {Jenkins}, {Borucki},
  {Bryson}, {Caldwell}, {Clarke}, {Dotson}, {Haas}, {Hall}, {Klaus}, {Koch},
  {McCauliff}, {Quintana}, {Twicken}, \& {van Cleve}}]{Gilliland2010}
{Gilliland}, R.~L., {Jenkins}, J.~M., {Borucki}, W.~J., {et~al.} 2010, \apjl,
  713, L160

\bibitem[{{Goldreich} \& {Kumar}(1988)}]{Goldreich1988}
{Goldreich}, P., \& {Kumar}, P. 1988, \apj, 326, 462

\bibitem[{{Hedman} \& {Nicholson}(2013)}]{Hedman2013}
{Hedman}, M.~M., \& {Nicholson}, P.~D. 2013, \aj, 146, 12

\bibitem[{{Hedman} \& {Nicholson}(2014)}]{Hedman2014}
---. 2014, \mnras, 444, 1369

\bibitem[{{Howell} {et~al.}(2014){Howell}, {Sobeck}, {Haas}, {Still},
  {Barclay}, {Mullally}, {Troeltzsch}, {Aigrain}, {Bryson}, {Caldwell},
  {Chaplin}, {Cochran}, {Huber}, {Marcy}, {Miglio}, {Najita}, {Smith},
  {Twicken}, \& {Fortney}}]{Howell2014}
{Howell}, S.~B., {Sobeck}, C., {Haas}, M., {et~al.} 2014, \pasp, 126, 398

\bibitem[{{Le Bihan} \& {Burrows}(2013)}]{LeBihan2013}
{Le Bihan}, B., \& {Burrows}, A. 2013, \apj, 764, 18

\bibitem[{{Leibacher} \& {Stein}(1981)}]{Leibacher1981}
{Leibacher}, J.~W., \& {Stein}, R.~F. 1981, NASA Special Publication, 450

\bibitem[{{Lissauer} {et~al.}(2014){Lissauer}, {Marcy}, {Bryson}, {Rowe},
  {Jontof-Hutter}, {Agol}, {Borucki}, {Carter}, {Ford}, {Gilliland}, {Kolbl},
  {Star}, {Steffen}, \& {Torres}}]{Lissauer2014}
{Lissauer}, J.~J., {Marcy}, G.~W., {Bryson}, S.~T., {et~al.} 2014, \apj, 784,
  44

\bibitem[{{Marley}(1991)}]{Marley1991}
{Marley}, M.~S. 1991, \icarus, 94, 420

\bibitem[{{Marley} \& {Porco}(1993)}]{Marley1993}
{Marley}, M.~S., \& {Porco}, C.~C. 1993, \icarus, 106, 508

\bibitem[{{More} {et~al.}(1980){More}, {Garbow}, \& {Hillstrom}}]{More1980}
{More}, J., {Garbow}, B., \& {Hillstrom}, K. 1980, Argoone National Laboratory
  Report ANL-80-74

\bibitem[{{Morton} {et~al.}(2016){Morton}, {Bryson}, {Coughlin}, {Rowe},
  {Ravichandran}, {Petigura}, {Haas}, \& {Batalha}}]{Morton2016}
{Morton}, T.~D., {Bryson}, S.~T., {Coughlin}, J.~L., {et~al.} 2016, \apj, 822,
  86

\bibitem[{{Mosser}(1990)}]{Mosser1990}
{Mosser}, B. 1990, \icarus, 87, 198

\bibitem[{{Mosser}(1995)}]{Mosser1995}
---. 1995, \aap, 293, 586

\bibitem[{{Mosser} \& {Appourchaux}(2009)}]{Mosser2009b}
{Mosser}, B., \& {Appourchaux}, T. 2009, \aap, 508, 877

\bibitem[{{Mosser} {et~al.}(2000){Mosser}, {Maillard}, \&
  {M{\'e}karnia}}]{Mosser2000}
{Mosser}, B., {Maillard}, J.~P., \& {M{\'e}karnia}, D. 2000, Icarus, 144, 104

\bibitem[{{Mosser} {et~al.}(1993){Mosser}, {Mekarnia}, {Maillard}, {Gay},
  {Gautier}, \& {Delache}}]{Mosser1993}
{Mosser}, B., {Mekarnia}, D., {Maillard}, J.~P., {et~al.} 1993, \aap, 267, 604

\bibitem[{{Mosser} {et~al.}(1996){Mosser}, {Galdemard}, {Lagage}, {Pantin},
  {Sauvage}, {Lognonne}, {Gautier}, {Billebaud}, {Livengood}, \&
  {Kaufl}}]{Mosser1996}
{Mosser}, B., {Galdemard}, P., {Lagage}, P., {et~al.} 1996, Icarus, 121, 331

\bibitem[{{Mosser} {et~al.}(2009){Mosser}, {Michel}, {Appourchaux}, {Barban},
  {Baudin}, {Boumier}, {Bruntt}, {Catala}, {Deheuvels}, {Garc{\'{\i}}a},
  {Gaulme}, {Regulo}, {Roxburgh}, {Samadi}, {Verner}, {Auvergne}, {Baglin},
  {Ballot}, {Benomar}, \& {Mathur}}]{Mosser2009c}
{Mosser}, B., {Michel}, E., {Appourchaux}, T., {et~al.} 2009, \aap, 506, 33

\bibitem[{{Mosser} {et~al.}(2010){Mosser}, {Belkacem}, {Goupil}, {Miglio},
  {Morel}, {Barban}, {Baudin}, {Hekker}, {Samadi}, {De Ridder}, {Weiss},
  {Auvergne}, \& {Baglin}}]{Mosser2010}
{Mosser}, B., {Belkacem}, K., {Goupil}, M.-J., {et~al.} 2010, \aap, 517, A22

\bibitem[{{Press} {et~al.}(1992){Press}, {Teukolsky}, {Vetterling}, \&
  {Flannery}}]{Press1992}
{Press}, W.~H., {Teukolsky}, S.~A., {Vetterling}, W.~T., \& {Flannery}, B.~P.
  1992, {Numerical recipes in FORTRAN. The art of scientific computing},
  418--423

\bibitem[{Rowe(2016)}]{jason_rowe_2016_60297}
Rowe, J. 2016, Kepler: Kepler Transit Model Codebase Release.,
  doi:10.5281/zenodo.60297

\bibitem[{{Rowe} {et~al.}(2006){Rowe}, {Matthews}, {Seager}, {Kuschnig},
  {Guenther}, {Moffat}, {Rucinski}, {Sasselov}, {Walker}, \&
  {Weiss}}]{Rowe2006}
{Rowe}, J.~F., {Matthews}, J.~M., {Seager}, S., {et~al.} 2006, \apj, 646, 1241

\bibitem[{{Rowe} {et~al.}(2008){Rowe}, {Matthews}, {Seager}, {Miller-Ricci},
  {Sasselov}, {Kuschnig}, {Guenther}, {Moffat}, {Rucinski}, {Walker}, \&
  {Weiss}}]{Rowe2008}
---. 2008, \apj, 689, 1345

\bibitem[{{Rowe} {et~al.}(2014){Rowe}, {Bryson}, {Marcy}, {Lissauer},
  {Jontof-Hutter}, {Mullally}, {Gilliland}, {Issacson}, {Ford}, {Howell},
  {Borucki}, {Haas}, {Huber}, {Steffen}, {Thompson}, {Quintana}, {Barclay},
  {Still}, {Fortney}, {Gautier}, {Hunter}, {Caldwell}, {Ciardi}, {Devore},
  {Cochran}, {Jenkins}, {Agol}, {Carter}, \& {Geary}}]{Rowe2014}
{Rowe}, J.~F., {Bryson}, S.~T., {Marcy}, G.~W., {et~al.} 2014, \apj, 784, 45

\bibitem[{{Roxburgh} \& {Vorontsov}(2006)}]{Roxburgh2006}
{Roxburgh}, I.~W., \& {Vorontsov}, S.~V. 2006, \mnras, 369, 1491

\bibitem[{{Schmider} {et~al.}(1991){Schmider}, {Fossat}, \&
  {Mosser}}]{Schmider1991}
{Schmider}, F.-X., {Fossat}, E., \& {Mosser}, B. 1991, \aap, 248, 281

\bibitem[{{Simon} {et~al.}(2016){Simon}, {Rowe}, {Gaulme}, {Hammel},
  {Casewell}, {Fortney}, {Gizis}, {Lissauer}, {Morales-Juberias}, {Orton},
  {Wong}, \& {Marley}}]{Simon2016}
{Simon}, A.~A., {Rowe}, J.~F., {Gaulme}, P., {et~al.} 2016, \apj, 817, 162

\bibitem[{{Stetson}(1987)}]{Stetson1987}
{Stetson}, P.~B. 1987, \pasp, 99, 191

\bibitem[{{Vorontsov} {et~al.}(1976){Vorontsov}, {Zharkov}, \&
  {Lubimov}}]{Vorontsov1976}
{Vorontsov}, S.~V., {Zharkov}, V.~N., \& {Lubimov}, V.~M. 1976, \icarus, 27,
  109

\bibitem[{{Walker} {et~al.}(2003){Walker}, {Matthews}, {Kuschnig}, {Johnson},
  {Rucinski}, {Pazder}, {Burley}, {Walker}, {Skaret}, {Zee}, {Grocott},
  {Carroll}, {Sinclair}, {Sturgeon}, \& {Harron}}]{Walker2003}
{Walker}, G., {Matthews}, J., {Kuschnig}, R., {et~al.} 2003, \pasp, 115, 1023

\bibitem[{{Walter} {et~al.}(1996){Walter}, {Marley}, {Hunten}, {Sprague},
  {Wells}, {Dayal}, {Hoffmann}, {Sykes}, {Deutsch}, {Fazio}, \&
  {Hora}}]{Walter1996}
{Walter}, C.~M., {Marley}, M.~S., {Hunten}, D.~M., {et~al.} 1996, \icarus, 121,
  341

\bibitem[{{Zhu} {et~al.}(1997){Zhu}, {Byrd}, \& {Nocedal}}]{Zhu1997}
{Zhu}, C., {Byrd}, R.~H., \& {Nocedal}, J. 1997, ACM Transactions on
  Mathematical Software, 23, 550

\end{thebibliography}

%\begin{figure}
%\begin{center}
%\includegraphics[scale=0.6]{timingpaperFigure.eps}
%\end{center}
%\caption[Geometry]{The geometry of two spatially separated bodies represented by yellow disks.  The top body represents a distant gravitationally bound binary system with the companion represented by a red disk placed at observed located along it's plots circular path.  The bottom body can represent the Sun, with the orbit of the Earth represented by the circle with radius, $a_o$.  The light-blue disks mark the locations of observers along the Earth's orbital path.  In this example the orbital plane of the distant binary is perfectly aligned along the Earth-Sun ecliptic.  We define $d$ as a distance from the observer to the central star at quadrature, $a_o$ as the distance from the observer to the Sun and $\phi$ as the angle between the observer and host star observed near quadrature (1) and near opposition (2).}
%\label{fig:geometry}
%\end{figure}

\begin{deluxetable}{ccc}
\tabletypesize{\scriptsize}
\tablecaption{Adopted Photometry}
\tablewidth{0pt}
\tablehead{
\colhead{Time}   & \colhead{Flux} & \colhead{Uncertainty}
}
\startdata
   16.40017500   &   -0.00317 & 0.00036 \\
   16.40085595   &   -0.00292 & 0.00036 \\
   16.40153699   &   -0.00269 & 0.00036 \\
   16.40221805   &   -0.00265 & 0.00036 \\
   16.40289918   &   -0.00265 & 0.00036
\enddata
\tablecomments{Adopted distance corrected photometry.  Reported time corresponds to BJD-2456977.10319595}
\label{tab:phot}
\end{deluxetable}

\begin{deluxetable}{ccc}
\tabletypesize{\scriptsize}
\tablecaption{Fitted Parameters}
\tablewidth{0pt}
\tablehead{
\colhead{Parameter}   & \colhead{Value} & \colhead{Units}
}
\startdata
$B$ & 2.25$\times10^{-2}$ & ppm$^2$/\uhz \\
$A_1$ & 8.32$\times10^{4}$ & ppm$^2$/\uhz \\ 
$\theta_1$ & 2.55 $\times10^{1}$ & 1/\uhz \\
$\alpha_1$ & 2.28 & \\
$A_2$ & 6.01$\times10^{4}$ & ppm$^2$/\uhz \\ 
$\theta_2$ & 3.63 $\times10^{-2}$ & 1/\uhz \\
$\alpha_2$ & 5.54 & \\
$D$ & 1.30 $\times10^{-2}$ & ppm$^2$/\uhz \\
$f_{max}$ & 3.25 $\times10^{3}$ & \uhz \\
$\sigma$ & 9.23 $\times10^{1}$ & \uhz
\enddata
\tablecomments{Best fit parameters.}
\label{tab:fitpars}
\end{deluxetable}

\listofchanges

\end{document}